\documentclass[doublecol]{epl2} 
\usepackage{amsmath}
\usepackage{amsfonts}

\newcommand{\ZZ}{\mathbb{Z}}

\title{Spectral analysis of finite-time correlation matrices near equilibrium phase transitions}
\shorttitle{Spec. ann. fin. corr. mat. equil. phase tran. } 

\author{Vinayak\inst{1} \and T.~Prosen\inst{2,3} \and B.~Bu\v ca\inst{2}\and T.~H.~Seligman\inst{2,3}}
\shortauthor{Vinayak \etal}

\institute{                    
  \inst{1} Instituto de Ciencias F\' isicas, Universidad Nacional Aut\' onoma de M\' exico, C.P. 62210 Cuernavaca, M\' exico\\
  \inst{2} Department of Physics, FMF, University of Ljubljana, Jadranska 19, 1000 Ljubljana, Slovenia\\
  \inst{3} Centro Internacional de Ciencias, C.P. 62210 Cuernavaca, M\' exico
}

\abstract{
We study spectral densities for systems on lattices, which, at a phase transition display, power-law spatial correlations. Constructing the spatial correlation matrix we prove that its eigenvalue density shows a power law that can be derived from the spatial correlations. In practice time series are short in the sense that they are either not stationary over long time intervals or that they are not available over long time intervals. Also we usually do not have time series for all variables available. We shall make numerical simulations on a 2-D Ising model with the usual Metropolis algorithm as time-evolution. Using all spins on a grid with periodic boundary conditions we find a power law, that is, for large grids, compatible with the analytic result. We still find a power law even if we choose a fairly small subset of grid points at random. The exponents of the power laws will be smaller under such circumstances. For very short time series leading to singular correlation matrices we use a recently developed technique to lift the degeneracy at zero in the spectrum and find a significant signature of critical behavior even in this case as compared to high temperature results which tend to those of random matrix models.}

\begin{document}

\maketitle

\section{Introduction}
Correlation matrices have attracted attention for almost a hundred years starting with multivariate analysis in finance \cite{bachelier} and biology \cite{Wishart}. The latter introduced random matrices as a null hypothesis at this very early stage. Surprisingly this tool was almost not used in physics and in particular in the description of phase transitions. This is all the more surprising, as correlations are at the very heart of practically any analysis of phase transitions. We shall show, that the correlation matrix and its eigenvalues are potentially very useful in the discussion of critical phenomena. For a discrete system at the critical temperature a power law tail in its spatial correlations on the grid is typical. We shall show that this implies a power law for the spectral density of the correlation matrix, and we give the relation for the exponents.

While this sounds nice{,} it is actually more difficult to detect then spatial correlations if we have the correlations for all the grid or even for a compact finite subset of points, that extends further {than} the localization length. Yet this is typically {\it not} available if we have experimental time series from a reduced set of points whose spatial distribution may not be known.  In order to emulate such a situation, as well as others mentioned below, we make numerical studies using the 2-D Ising model with standard Metropolis dynamics as a paradigmatic example. We first reproduce the situation with long time series and the full correlation matrix and obtain a power law with an exponent compatible with the analytic result. This is in clear contrast to {the random matrix theory (RMT) result, namely} the Mar\v cenko Pastur distribution {\cite{marchenko, Mehta,Brody81,TGW:98,vp2010}}, which unsurprisingly appears at high temperatures. Next we randomly ch{o}ose a subset of grid points that may represent a rather small fraction of the total and whose spatial relation is not used. We find very satisfactorily that the power law persists though with a slightly smaller exponent near the critical temperature{. Yet if the number of eigenvalues gets too small, the range of the power law shrinks to the point, where we will need other indicators of correlations due to criticality or other reasons.}

{Indeed} in a practical context we often have to deal with {very} short time series, particularly if stationarity limits the time horizon. By short we imply, that the number of time series used to construct the correlation matrix is larger than the length of the time series. This leads, due to their dyadic product structure, to singular correlation matrices. The number of nonzero eigenvalues is determined by the length of the time series, and if we increase the number {of time series} and thus the size of the correlation matrix the number of {nonzero} eigenvalues will not change. {W}e believe that the larger matrix does contain more information. {Collecting more data for additional time series might be expensive, and thus we have to know if there is more information and if so, how to exploit it. Such} information can be retrieved by considering the entire matrix, as indicated in a financial market context in Ref. \cite{Thomas2012}; see also Refs. \cite{Bio:1, Bio:2, markus, Ruben} in the context of biology. Yet this is cumbersome due to the large number of matrix elements, and we would like to limit our considerations to the analysis of eigenvalues {\cite{Bouchaud:2009,StanleyGuhr:2002, Stanley:1999, Potter:1999}}. We propose to use a general method {\cite{vrt2013}} that seems productive and sensitive to the question whether we have correlations among our time series or not. Using the power map as introduced in Refs. \cite{GuhrKabler2003,GuhrShafer2010} with the original purpose of reducing noise, in Ref. \cite{vrt2013} it was shown that this map is effective in breaking the degeneracy at zero eigenvalue and thus lifting the singularity of the matrix{; for a pedagogical introduction see Ref. \cite{ELAF}}. With very small perturbations, i.e. with powers sufficiently near to $1$ the spectrum so obtained is well separated from the original spectrum and has been termed {\it emerging spectrum}. In the same paper it has been shown for random matrix models, that the emerging spectrum is {more sensitive to correlations than the original or bulk spectrum}. {T}he same holds true {to a lesser degree} for the original spectrum, the emerging spectrum can {also} hold many more eigenvalues, and thus present better statistics if the number of time series $N$ is much larger than the length of the time series $\tau$. This analysis will allow us to have sensitive tools that not only serve to study the phase transition on hand, but hold promise to be useful in other cases, such as non-equilibrium phase transitions, say in stationary systems or bifurcations, e.g. saddle node or pitchfork bifurcations, in non-linear dynamical systems.

\section{ Scaling of the correlation spectrum at the critical point}
First we shall provide a simple analytical calculation of the spectrum of correlation matrix at the critical temperature. Suppose we have a system on a regular cubic $d-$dimensional lattice of size $L$, $\ZZ^d_L$, and assume that for asymptotically long times of observation $\tau\to\infty$, the correlations decay with exponent ${\theta} > 0$,
\begin{equation}
C_{\vec{n},\vec{m}} \approx \frac{c}{|\vec{n}-\vec{m}|^{{\theta}}},{\rm \quad for\quad}1 \ll |\vec{n}-\vec{m}| \ll L.
\label{scaling}
\end{equation}
For the Ising model on a square lattice $d=2$ at the critical temperature we have for example ${\theta}=1/4$. If, in addition, periodic boundary conditions are assumed, then $C_{\vec{n},\vec{m}}$ is a $d$-dimensional {\em circulant} matrix, i.e.  $C_{\vec{n},\vec{m}}= f(\vec{n}-\vec{m})$, where $f(\vec{n})$ is a function on a $\ZZ_L^d$. Therefore, $C_{\vec{n},\vec{m}}$ can be diagonalized in terms of a $d-$dimensional {\em discrete Fourier transformation}, yielding eigenvalues $\lambda^{(\vec{k})}$ and eigenfunctions $u_{\vec{n}}^{(\vec{k})}$, labeled again by points $\vec{k}$
on $\ZZ^d_L$.
\begin{equation}
\sum_{\vec{m}} f(\vec{n}-\vec{m}) u_{\vec{m}}^{(\vec{k})} = \lambda^{(\vec{k})} u_{\vec{n}}^{(\vec{k})}.
\label{eigen}
\end{equation}
As the left-hand-side of eigenvalue equation (\ref{eigen}) is a convolution,  we obtain $\lambda^{(\vec{k})}$ by transforming to momentum space $u_{\vec{n}}^{(\vec{k})} = \sum_{\vec{l}} v^{(\vec{k})}_{\vec{l}} e^{-i \vec{l}\cdot\vec{n}}$, namely
\begin{equation}
\lambda^{(\vec{k})} = \sum_{\vec{n}} f(\vec{n})e^{i \vec{k}\cdot \vec{n}},
\end{equation}
and $v^{(\vec{k})}_{\vec{l}} = \delta_{\vec{k},\vec{l}}$. For large $L\gg 1$, and given the asymptotic power-law scaling (\ref{scaling}) of $f(r) \approx c r^{-{\theta}}$, for $1 \ll r \ll L$, we can estimate the correlation eigenvalues for $1 \ll |\vec{k}| \ll L$ as
\begin{equation}
\lambda^{(\vec{k})} \approx \int {\rm d}^d \vec{r} \frac{c}{r^{\theta}} e^{i \vec{k}\cdot \vec{r}} = \frac{c'_d}{|\vec{k}|^{d-{\theta}}},
\label{int}
\end{equation}
where $c'_d$ is a constant which in principle depends only on the dimensionality $d$ and constant $c$. The last identity in (\ref{int}) simply follows from non-dimensionalizing the integral.
Labeling the eigenvalues by {\em decreasing eigenvalue} $\lambda_j$, $\lambda_1 \ge \lambda_2 \ge \lambda_3 \cdots$, we find that $j \propto |\vec{k}|^d$, and
\begin{equation}\label{gasmp}
\lambda_j \approx \frac{c_d}{j^{\zeta}}, \quad {\zeta}=\frac{d-{\theta}}{{d}},\quad{\rm for}\quad 1\ll j \ll N=L^d.
\end{equation}
therefore the following scaling should hold for correlation eigenvalues at the critical temperature. For example, for 2d Ising model we have ${\zeta} = \frac{7}{8}$.
\begin{figure}
        \centering
               \includegraphics [width=0.48\textwidth]{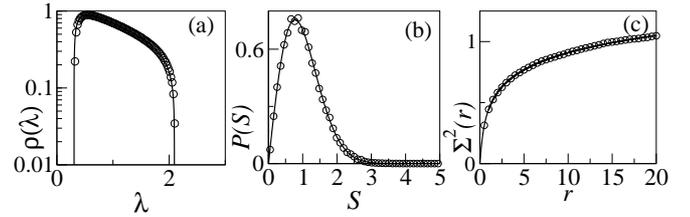}
                \caption{Spectral statistics of $\mathsf{C}$ at a high temperature, $2J/\mathcal{T}=0.001$, for $L=192$. In (a) we compare the spectral density with the Mar\'cenko Pastur formula. In (b) and (c), we compare the nearest-neighbor-spacing distribution, $P(S)$, and the number variance, $\Sigma^{2}(r)$, with those of RMT. Numerics are shown by open circles and {RMT results} are shown by solid lines.}
\label{mp1}
\end{figure}

\section{Numerics for the 2-dimensional Ising model}
To fix definitions and notation we shall give the Ising Hamiltonian {\cite{Khatun}} with {the} coupling strength $J$
{
\begin{equation} 
H_{\rm Ising} = -J\sum_{\langle i, j\rangle} \sigma_{i} \sigma_{j},
\end{equation}
where the spin variable $\sigma_{k}$ takes values $\pm 1$ at the $k$'th lattice site from the set of $L^{2}$ sites, and $\sum_{\langle i, j\rangle}$ represents summation over all distinct nearest-neighbor pairs of lattice sites. Next we} describe the Metropolis dynamics for this system: We shall use a 2-dimensional square grid of size $L \times L $ with periodic boundary conditions and choose the sites on this grid at random to perform a spin flip and check the resulting energy ${\mathcal{U}}$ using the Hamiltonian. If the energy is lowered we accept the change and if it is raised we accept the change with probability ${\exp[(\Delta \mathcal{U})/\mathcal{T}]}$ where $\mathcal{T}$ {is} the temperature at which the dynamics occur. The critical temperature for this system is $\mathcal{T}_{c}= 2J/{\text{ln}}(\sqrt{2}+1)$. One time step will be given by ${10} L^2$ flips. Correspondingly we denote by $\tau$ the length of the time-series (in number of time steps), and by  $\tau_{\ll}$ numbers of time steps that are much smaller than the dimension of the correlation matrix. We shall first concentrate on the case $\tau \gg L^2$, and analyze two situations: First use the complete $L^2 \times L^2 = N\times N$ correlation matrix, where we assume that we know all possible data of the problem. This is fine for our model, but not practical in any situation of observation or experiment. We therefore shall also look, by way of example, at a situation where only a quarter of the possible measurements are performed, i.e. the size of the correlation matrix is $N /4 \times N /4$, in order to see, if incomplete sets of time series {chosen at random} render the correlation matrix useless. 

\begin{figure}
                \onefigure[width=0.48\textwidth]{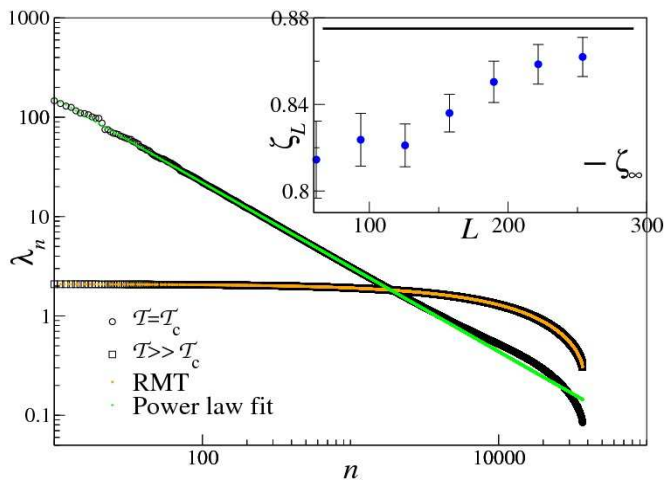}
                \caption{(Color online) The Zipf plot for the eigenvalues, $\lambda_{n}$, vs, the eigenvalue index, $n$, of $N\times N$ correlation matrices $\mathsf{C}$ at a very high temperature, as chosen in fig. \ref{mp1}, and at the critical temperature $\mathcal{T}_{c}$ for a square lattice of size $L=192$ where the time horizon $\tau=5N$. Data at $\mathcal{T}=\mathcal{T}_{c}$ are shown with {\it circles} while the high temperature data are shown with {\it squares}. The {{\it green} symbols} represent the power law fit where the exponent ${\zeta}=0.8504$. As expected, the high temperature plot coincides with {RMT} shown {in \it orange}. In the inset we show behavior of the exponent ${\zeta}$ {as a function of} the lattice dimension $L$. The solid black line is the asymptotic value of the exponent obtained from our theory (\ref{gasmp}). }
\label{PL-192}
\end{figure}

We start by looking at the spectral statistics for the first case at a very high temperature, viz. when $2J/\mathcal{T}=0.001$, for which {we} expect the predictions of RMT {to} hold. As shown in fig. \ref{mp1} the spectral density, $\rho(\lambda)${,} of eigenvalues $\lambda$ is closely described by the Mar\v cenko-Pastur result \cite{marchenko, vrt2013}. The fluctuations measures \cite{Mehta,Brody81,TGW:98}, e.g. the nearest-neighbor spacing distribution, $P(S)$, and the number variance, $\Sigma^{2}(r)$, taken for spectra normalized to average spacing one, coincide with those of the RMT \cite{Mehta,vp2010}. For the second case, i.e. the smaller correlation matrices results at this temperature show a similar agreement with RMT; we do not show the corresponding figures.

Next we turn to the central point, namely to the behavior of our system at critical temperature. We shall concentrate on the spectral one-point function, both because we have a theoretical result for this case and because the treatment of the two point functions causes significant finite size and finite time horizon effects. While we can use the density of states, the so called Zipf plot, i.e. a plot of the eigenvalues, ${\lambda_{n}}$, against the label, ${n}$, in an ordered spectrum seems to give more direct information on the power laws that may occur. In fig. \ref{PL-192} we show the Zipf plot for the eigenvalues of the $N\times N$ correlation matrix $\mathsf{C}$ at the critical temperature $\mathcal{T}_{c}$ and at the high temperature corresponding to results of fig. \ref{mp1}. At both temperatures we consider all sites of the lattice of size $L=192$ with the time horizon $\tau=5N$. At $\mathcal{T}=\mathcal{T}_{c}$ we see a rather nice power law behavior with an averaged value, ${\zeta}=0.8504$, for a range from $100<n<1000$. The power we extract is not identical to the theoretical value $0.875$ we obtained above, but it depends on the size of the lattice. In the inset of the fig. \ref{PL-192} we show, how it behaves as a function of the lattice size. We see a tendency compatible with the asymptotic result, yet the data do not allow a precise extrapolation. Note that the high temperature result follows RMT very well, and thus finite size effects are not important, as we would expect for a matrix of this size. {The RMT result we have used in fig. \ref{PL-192} is obtained numerically from the Mar\v cenko-Pastur distribution, as $n(\lambda)=N\int^{\lambda}\rho(x)dx$.}

\begin{figure}
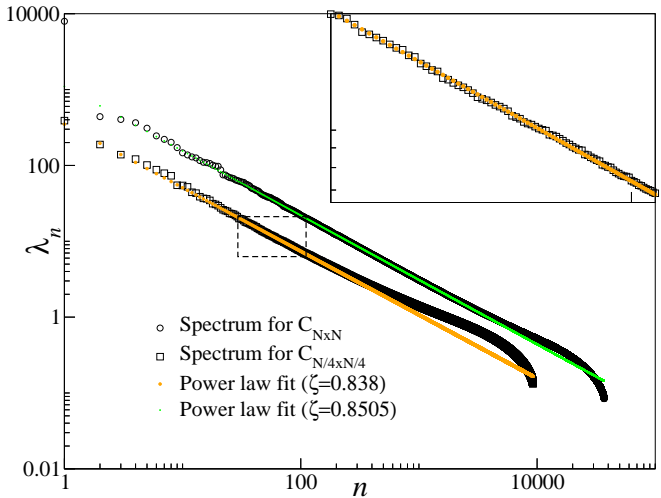

                \onefigure[width=0.48\textwidth]{NsmallN.eps}
                \caption{(Color online) The Zipf plot at the critical temperature $\mathcal{T}_{c}$. Here {\it black} open circles represent eigenvalues for {an} $N\times N$ matrix {where $N=L^{2}$ and $L=192$}, and {squares} represent eigenvalues for {an} $N/4\times N/4$ matrix {where the time series are chosen at random}. {In both cases,} $\tau$ is $5$ times the matrix dimension. The {{\it orange} and {\it green} symbols} are fits for the power law.}
\label{smallN}
\end{figure}
We now pass to the case of incomplete correlation matrices. In fig. \ref{smallN} we show the Zipf plot for the case the $N/4 \times N/4$ correlation matrix. In this case we still get a power law but at ${\zeta}=0.838$ for a range from $20<n<110$ {and} the difference to the asymptotic value is bigger. Nevertheless the power law is still very visible and the difference  to the high temperature results is notorious.

Note that the smaller eigenvalues at $\mathcal{T}_{c}$, in both figures, figs. \ref{PL-192} and \ref{smallN}, deviate from the power law and behave somewhat similar to the Mar\v cenko Pastur result we saw in the results for the high temperature. This behavior is more prominent in fig. \ref{smallN}, alluding to the randomness involved due to the finite number of sites.

\section{Singular correlation matrices and the emerging spectrum near criticality}
We now turn our attention to the case of short time series. Short here is defined as the case where the dyadic correlation matrix $\mathsf{C}$ has a dimension $D$ larger than the length of the time series $\tau_{ \ll} < D$.  Then $\mathsf{C}$ will have $\tau_{ \ll}$ {nonzero} eigenvalues and $D-\tau_{ \ll}$ eigenvalues zero. Particularly if $\tau_{\ll} \ll D$ , we will have a spectrum with few points and thus little chance to see a power law. In a recent paper the power map  \cite{GuhrKabler2003,GuhrShafer2010, vrt2013} was proposed to break the degeneracy of such spectrum at zero.

The power map is defined for the matrix elements of $\mathsf{C}$ as
\begin{eqnarray}\label{pmap}
C^{(q)}_{mn}&=&{\rm sgn}\,[C_{mn}] \, \big| C_{mn} \big|^q  \\
&=&C_{mn}\exp\left[\frac{(q-1)}{2}{\rm ln}{(C_{mn})}^{2}\right],
\end{eqnarray}
such that, for $q=1$ we retrieve the original correlation matrix. The non-linearity of the power map breaks the dyadic structure of $\mathsf{C}$ and thus, for $\tau_{ \ll}<D$, lifts the degeneracy of eigenvalues at $0$. For small $q-1$ the spectrum emerged from the $0$ eigenvalues (referred to henceforth as the { \it emerging spectrum}) is well separated from {t}he original {nonzero} spectrum, which we shall call the bulk. For RMT, where corrections are absent for $n\ne m$, the mean of the eigenvalues of the emerging spectrum is of order $O(q-1)$ and the mean of the corrections in the spectrum will be of the same order but with a negative sign \cite{vrt2013}. For $\tau{\sim} D/8$, the eigenvalues of the emerging spectra are positive. In contrast for the correlated case, even if $\tau\sim D/4$ negative eigenvalues will appear.

\begin{figure}
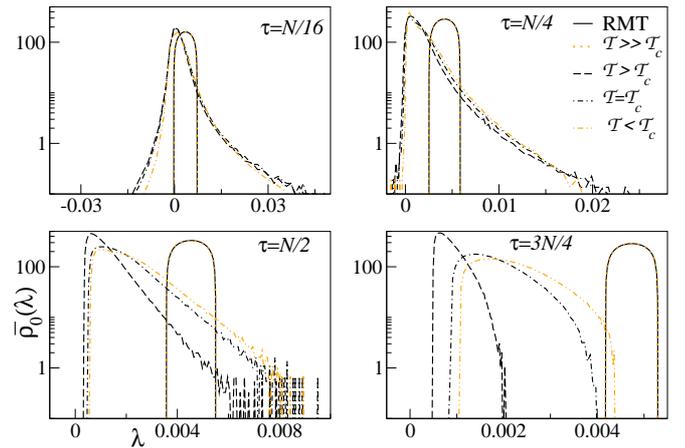

                \onefigure[width=0.48\textwidth]{Nden_emerg_log.eps}
                \caption{(Color online) Density of the emerging spectra, $\overline{\rho}_{0}(\lambda)$, under small-power map deformation where $q=1.001$. In this figure densities are compared near the critical temperature and at a very high temperature with those of RMT shown by solid {\it black} lines. Results are shown, in clockwise direction, respectively for $\tau=N/16,\,N/4,\,{3N/4}$, and ${N/2}$. {Dots, dashed lines, dashed-dotted lines, and dashed-double-dotted lines represent data respectively at high temperature, above critical temperature, critical temperature, and below critical temperature}.}
\label{den_emerg}
\end{figure}

This is seen in fig. \ref{den_emerg}, where we show the density of the emerging spectrum, $\overline{\rho}_{0}(\lambda)$, for different but short time horizons at several temperatures as indicated in the figure. For this figure we consider $L=192$ and construct $\mathsf{C}$ using all sites of the lattice. We first compare these densities at a very high temperature, viz. $2J/\mathcal{T}=0.001$. They coincide so closely with the RMT result that they become near indistinguishable.{ We remark that the RMT results we show in this figure are numerical. The first moment of this distribution is analytically known, as its the second moment for $\tau/N > 1/4$ \cite{vrt2013}.} Next, we compare densities near the critical temperature, ranging from $2J/\mathcal{T}=2J/\mathcal{T}_{c}-0.01$ to $2J/\mathcal{T}=2J/\mathcal{T}_{c}+0.01$. These results differ strongly from the RMT case, both in the shape of the density of eigenvalues and in the emergence of negative eigenvalues as we diminish the size of {$\tau$}. Note that for {the} larger {$\tau$} sensitivity to temperature near criticality is bigger than for smaller ones.
\section{Conclusion} Summarizing, we have three central points. First  we find a derived power law in the spectrum of the correlation matrix for systems with power law spatial correlations. While this is not entirely surprising we have found no previous use of the eigenvalues of the correlation matrices of critical periodic systems. {Note that the nonzero eigenvalues of the temporal and spatial correlation matrices are identical. Thus the derived power law for these eigenvalues is the same in space and time. Cross correlations in time may nevertheless be useful objects of study. Time delayed correlation matrices can then be constructed; this gives rise to non-symmetric correlation matrices \cite{Bouchaud:2009,Drodz, vin2013,vinLuis2014} whose singular values can be studied}. Next, we use the two-dimensional Ising model with Metropolis dynamics as the simplest example for numerical studies. There we find that restricting the number of time series to an randomly chosen small subset without any order with respect to the spatial arrangement, we still find a power law, which is indeed similar to the original one. Finally we take into account that in practice we often have to use short time series as stationarity of the processes involved may not be guaranteed over long time horizons. Near the critical temperature distinctive critical behavior can also be seen with short time series. At high temperatures, unsurprisingly we find universal random matrix behavior. The distinctive behavior can best be seen if we use techniques we recently developed, to show in short time series that lift the degeneracy at zero in the spectrum.  We demonstrate high sensitivity to the correlations present in critical or near critical cases and as the spatial correlations induce dramatic changes in what we call the emerging spectrum, which results from lifting the degeneracy. In this way we obtain more, or at least statistically better spectral information as we use a larger number of time series, i.e. as more data are made available with the same short time horizon, despite of the resulting singularity of the original spectrum.

We have thus not only found interesting behavior of the correlation matrix of critical periodic systems, but we have also used these features to demonstrate the power of new tools to analyze large sets of short time series typical of for quasi-stationary systems.




\acknowledgments
We are grateful to M. \v Znidari\v c, T. Guhr, F. Leyvraz and P. Zanardi for useful discussions. We acknowledge financial support by the program P1-0044 of the Slovenian Research Agency as well as by  UNAM/DGAPA/PAPIIT, grant {IN114014} and CONACyT, grant 154586. One of us (T.H.S.) is acknowledging a fellowship for distinguished scholars by the Slovenian Research Agency. Vinayak was supported by DGAPA/UNAM as a postdoctoral fellow. The heavier calculations were performed on UNAM's supercomputer MIXTLI.


\begin{thebibliography}{0}
\bibitem{bachelier}
  \Name{ Bachelier~L.}
  \Book{ The Random Character of Stock Market Prices}
  \Publ{ Gauthier-Villars, Paris}
  \Year{1964}
  \Page{17}.


\bibitem{Wishart}
  \Name{Wishart~J.}
  \REVIEW{ Biometrika}{20A}{1928}{32}.

\bibitem{marchenko}
  \Name{Mar\v cenko~V.~A. \and  Pastur~L.~A.}
  \REVIEW{Math.~USSR~Sb.}{1}{1967}{457}.

\bibitem{Mehta}
  \Name{Mehta~M.~L.}
  \Book{  Random Matrices}
  \Publ{Academic Press, New York}
  \Year{ 2004}.

\bibitem{Brody81}
  \Name{ Brody~T.~A., Flores~J., French~J.~B., Mello~P.~A., Pandey~A. \and Wong~S.~S.~M.}
  \REVIEW{ Rev. Mod. Phys.}{53}{1981}{385}.

\bibitem{TGW:98}
  \Name{Guhr~T., Groeling~A.~M.\and  Weidenm\"{u}ller~H.~A.}
  \REVIEW{Phys. Rep.}{299}{1998}{189}.

\bibitem{vp2010}
  \Name{Vinayak \and Pandey~A.}
  \REVIEW{ Phys. Rev. E}{81}{2010}{036202}.

\bibitem{Thomas2012}
  \Name{M\"{u}nnix~M.~C., Shimada~T.,  Sch\"{a}fer~R., Leyvraz~F., Seligman~T.~H., Guhr~T., \and  Stanley~H.~E.}
  \REVIEW{ Sci. Rep.}{2}{2012}{644}.

{
\bibitem{Bio:1}
 \Name{Lee~I., Yoganarasimha~D., Rao~G., \and Knierim~J.~J.}
 \REVIEW{Nature}{430}{2004}{456}.
\bibitem{Bio:2}
 \Name{Johansen-Berg~H., Behrens~ T.~E.~J., Robson~M.~D., Drobnjak~I.,
Rushworth~M.~F.~S., Brady~J.~M., Smith~S.~M., Higham~D.~J. \and Matthews~P.~M.}
 \REVIEW{ Proc. Natl. Acad. Sci. USA}{101}{2004}{13335}.
\bibitem{markus}
 \Name{M\"{u}ller~M., Baier~G., Galka~A., Stephani~U. \and Muhle~H.}
 \REVIEW{Phys. Rev. E }{71}{2005}{046116}.
\bibitem{Ruben}
 \Name{Fossion~R.}
 \REVIEW{AIP Conf. Proc.}{1575}{2014}{89}.

\bibitem{Bouchaud:2009}
 \Name{Bouchaud~J.~P. \and Potters~M.}
  \REVIEW{arXiv}{q-fin.ST}{2009}{0910.1205}.
\bibitem{StanleyGuhr:2002}
 \Name{Plerou~V., Gopikrishnan~P., Rosenow~B., Amaral~L.~A.~N., Guhr.~T., \and Stanley~H.~E.}
 \REVIEW{Phys. Rev. E.}{65}{2002}{066126}.
\bibitem{Stanley:1999} 
 \Name{Plerou~V., Gopikrishnan~P., Rosenow~B., Amaral~L.~A.~N., \and Stanley~H.~E.}
 \REVIEW{Phys. Rev. Lett.} {83}{1999}{1471}.
\bibitem{Potter:1999}
 \Name{Laloux~L., Cizeau~P., Bouchaud~J.~P., \and Potters~M.}
 \REVIEW{Phys. Rev. Lett.}{83}{1999}{1467}.
}

\bibitem{vrt2013}
  \Name{Vinayak, Sch\"{a}fer~R. \and Seligman~T.~H.}
  \REVIEW{Phys. Rev. E}{88}{2013}{032115}.



\bibitem{GuhrKabler2003}
  \Name{Guhr~T.\and K\"{a}lber~B.}
  \REVIEW{ J. Phys. A: Math.Gen.}{36}{2003}{3009}.

\bibitem{GuhrShafer2010}
  \Name{Sch\"{a}fer~R., Nilsson~N.~F. \and Guhr~T.}
  \REVIEW{Quantitative Finance}{10}{2010}{107}.


{
\bibitem{ELAF}
  \Name{Vinayak \and Seligman~T.~H.}
  \REVIEW{AIP Conf. Proc.}{1575}{2014}{196}. 
\bibitem{Khatun}
  \Name{Khatun~M., Barry~J.~H. \and Tanaka~T.}
   \REVIEW{Phys. Rev. B}{42}{1990}{4398}.

\bibitem{vin2013}
 \Name{Vinayak}
  \REVIEW{Phys. Rev. E}{88}{2013}{042130}.

\bibitem{vinLuis2014}
 \Name{Vinayak \and Benet~L.}
  \REVIEW{arXiv}{math-ph}{2014}{1403.7250}.

\bibitem{Drodz}
 \Name{Kwapie\'n~J., Dro\.zd\.z~S.}
  \REVIEW{Phys. Rep.}{515}{2012}{115}.
}

\end{thebibliography}
\end{document}